\documentclass[prd,aps]{revtex4}

\usepackage{graphicx,subfigure}

\newcommand{\be}{\begin{equation}}
\newcommand{\ee}{\end{equation}}
\newcommand{\bea}{\begin{eqnarray}}
\newcommand{\eea}{\end{eqnarray}}
\newcommand{\bd}{\begin{displaymath}}
\newcommand{\ed}{\end{displaymath}}

\newcommand{\lb }{ \left( }
\newcommand{\rb }{ \right ) }
\newcommand{\qw }{ D_{q,w} }

\begin{document}

%\draft
%\preprint{
%\begin{tabular}{l}
%\hbox to\hsize{\hfill KAIST-TH 2006/10}\\
%[-1mm]
%\hbox to\hsize{\hfill KIAS-P06047}\\
%[-2mm] \hbox to\hsize{\hfill hep-ph/yymmdd}\\
%[-3mm] \hbox to\hsize{\hfill November 2011}\\
%[-3mm]
%\end{tabular}
%}

\title{
Study on the mechanical system related to Hahn's discrete time derivative
}

\author{ Won Sang Chung }
\email{mimip4444@hanmail.net}

\affiliation{
Department of Physics and Research Institute of Natural Science, College of Natural Science, Gyeongsang National University, Jinju 660-701, Korea
}

\author{ Min Jung }
\email{sosost1004@nate.com}

\affiliation{
Department of Physics and Research Institute of Natural Science, College of Natural Science, Gyeongsang National University, Jinju 660-701, Korea
}

\date{\today}

\begin{abstract}
In this paper, we use the quantum variational calculus related to Hahn's discrete time derivative construct the deformed version for the classical mechanics related to the Hahn's calculus. We deal with the deformed dynamics such as the motion with constant velocity and the motion with constant acceleration. Moreover, we extend our work to the motion of a body in a resisting medium by using the new identity for an infinite ($q,w$)-series , where the retarding force is assumed to be proportional to the deformed average velocity.

\end{abstract}

%\pacs{PACS numbers:12.60.Fr,12.60.Cn,14.80.Cp}
\maketitle

\section{Introduction}

Quantum difference operators are receiving an increasing interest in the applied mathematics and the theoretical physics because of their application [1-5]. Briefly Speaking, a quantum calculus substitute the ordinary derivative by a difference operator, which allows us to treat sets of non-differentiable functions. Since Jackson [6] introduced the first expression of difference operator called a Jackson derivative, several expressions of the difference operator appeared. Among them, the most famous one is Hahn's difference operator [7], which has two deformation parameters $w$ and $q$. Hahn's operator becomes Jackson's q-derivative when the parameter $w$ goes to $0$.

Recently, Malinowska and Torres [8] studied the Hahn quantum variational calculus, in which they studied the Hahn quantum Lagrange problem. From now on, we will restrict our concern to the case that the parameter $w$ goes to $0$. Indeed, their action has the following form :
\be
{ \cal S } [x] = \int_a^b {\cal L}( x, D_t x ; t ) d_q t,
\ee
where $x$ satisfies some boundary condition. From their quantum variation calculus, we can obtain the following equation minimizing the action given in the eq.(1) :
\be
D_t ( \frac { \partial   {\cal L} } { \partial D_t x } ) -  D_x {\cal L}   =0
\ee

Hahn's discrete derivative is defined as
\be
\qw f(t) = \cases { \frac { f(qt + w ) -f(t) }{(q-1)t + w } & ($t \ne w_0 $)\cr
f'(t) & ( $t = w_0 $) \cr },
\ee
where $ w_0 = \frac {\tau }{ 1-q} $

The Leibnitz rule for Hahn derivative is as follows :
\bd
\qw ( f(t)g(t) ) = ( \qw f(t) ) g(t) + f ( qt + w ) \qw g(t)
\ed
\be
\qw \lb \frac { f(t) }{g(t) } \rb = \frac { ( \qw f(t) ) g(t) - f ( t ) \qw g(t)}{ g(t) g( qt+w) }
\ee
The Hahn integral is defined as the inverse operation of the Hahn derivative :
\be
\int_{w_0 }^ t f(t') d_{q,w}t' = ( (1-q)t-w) \sum_{k=0}^{\infty} q^k f( tq^k + [k]_{q,w} ),
\ee
where the ($q,w$)-deformed number is defined as
\be
[k]_{q,w}= \frac {w ( 1-q^k )}{1-q}
\ee
The $q,w$-deformed number is proportional to the Jackson's $q$-deformed number as follows :

\be
[k]_{q,w}= w [k]_q ,
\ee
where
\be
[k]_q  = \frac {w ( 1-q^k )}{1-q}
\ee

We have the following useful formula from the definition of the Hahn's derivative :
\be
\qw (at+b)^n = a   \sum_{k=0}^{n-1} ( a(q t +w ) +b )^k (at+b)^{n-k-1}
\ee

Inserting $ a=1, b=0 $ into the eq.(9) gives

\be
\qw t^n =    \sum_{k=0}^{n-1} ( q t +w )^k t^{n-k-1}
\ee

Now let us introduce the following polynomial of degree $n$ as follows :
\be
( t; q, w )_n = \prod_{j=1}^n ( t- [j]_{q,w} )
\ee

Then we have the following property:
\be
\qw (t;q,w)_n = [n]_{q,0} ( t; q,w)_{n-1}
\ee

This work urges us to construct the deformed classical mechanics related to Hahn's calculus, which is the main purpose of this paper.

\section { ($q,w$)-analogue of Galilei's formula}

With the formulation of the Hahn's calculus, we are able to construct the deformation of the classical mechanics. Let us introduce the deformed velocity $v(t)$and deformed acceleration $a(t)$ as follows :
\be
v(t) = D_t x(t) , ~~~~ a(t) = D_t v(t),
\ee
where $x(t)$ is a deformed position and the deformed time derivative $D_t $ is a Hahn's derivative defined by :
\be
D_t f(t) = \frac { f(qt+w) - f(t) }{t(q-1)+w }
\ee
The eq.(13) is easily derived form the deformed Lagrangian defined in the eq.(1).
We restrict our concern to the case that the $ w$ is real and $q$ satisfies $ 0<q < 1 $. When $ w$ goes to $0$ and $q$ goes to 1, a deformed time derivative reduces to an ordinary time derivative.

Now let us start with the motion with the constant velocity $v$. The relation between the deformed position and the deformed velocity as follows :
\be
D_t x = \frac { x(qt+w) - x(t) }{t(q-1) +w} = v
\ee
It can be written as
\be
x(qt+w) - x(t) = (q-1) vt + v w
\ee
Iterating the eq.(16) $N$ times, we obtain
\be
x(q^N t + [N]_q w) - x(t) = (q-1) [N]_q  vt + [N]_q w v ,
\ee
where we used the following formulas :
\be
q^k ( qt + w ) + [k]_q w = q^{ k+1} t + [ k+1]_q w
\ee
and
\be
\sum_{k=0}^{N-1} [k]_q = \frac {1}{1-q} ( N- [N]_q )
\ee
Inserting $ t=0 $ , we get
\be
x(w) - x(0) = v w
\ee
Now let us take the limit $ N \rightarrow \infty $ in the eq.(17). Then the eq.(17) becomes
\be
x(t) = x(w_0 ) + vt - vw_0
\ee
Inserting $t=0$ into the eq.(21), we get
\be
x(o) = x(w_0 ) - vw_0
\ee
and
\be
x(t) = x(0 ) + vt
\ee
Indeed, the solution (23) satisfies the relation (16).

Now let us discuss the motion with the constant acceleration $a$. The relation between the deformed velocity and the constant acceleration is given by
\be
\frac { v(qt+w) - v(t) }{t(q-1)+w } = a  ~~~or ~~~v(qt+w) - v(t) = (q-1) at + a w
\ee
It gives the following simple relation :
\be
v(t) = v(0) + a t
\ee
Because the deformed acceleration is given by applying the Hahn's derivative to the deformed velocity, we have
\be
D_t x(t) = v(0) + a t
\ee
or
\be
x(qt+w ) - x(t) = ( t(q-1) + w ) ( v(0) + a t)
\ee
Iterating the relation (27) $N$ times, we have
\be
x( q^N t + [N]_q w ) - x(t)
= \sum_{k=0} ^{N-1} q^k ( t(q-1) + w ) ( v(0) + a ( q^k t + [k]_q  w ) )
\ee
Using the following property
\be
\sum_{k=0}^{N-1} q^k [k]_q = \frac {q}{q+1}  [N]_q [N - 1 ]_q ,
\ee
we obtain the following relation from the eq.(28) :
\be
x( q^N t + [N]_q w ) - x(t)
= ( t(q-1) + w ) ( v(0) [N]_q + a t [N]_{q^2 } + \frac {a w}{ q +1 } [N]_q [N - 1 ]_q ) ,
\ee
Taking the limit $ N \rightarrow \infty $, the eq.(30) is then given by
\be
x(t) = x(w_0) - v(0) w_0 - \frac { a w_0^2 } {q+1} + v(0) t + \frac {a} { q +1 } t^2
\ee
Inserting $ t=0$ into the eq.(31), we have
\be
x(0) = x(w_0) - v(0) w_0 - \frac { a w_0^2 } {q+1},
\ee
which gives the following solution :
\be
x(t) = x(0) + v(0) t + \frac {1}{[2]_q } at^2
\ee

\section { Galilei's formula by using the second order ($q,w$)-difference equation.}

Because the deformed acceleration is given by applying the Hahn's derivative to the deformed position twice , we obtain the following relation :
\be
D_t^2 x(t) = a
\ee
From the definition of the deformed time derivative, we can obtain the following second order difference equation :
\be
x(q^2 t + [2]_q w) - [2]_q x(qt+q) + q x(t) = q a ( t(q-1) + w )^2
\ee
If we set
\be
h(t) = x(qt+w ) - q x(t),
\ee
we obtain
\be
h(qt+w ) - h(t) = q a ( t(q-1) + w )^2
\ee
Iterating the relation (37) $N$ times, we get
\be
h(q^N t + [N]_q w ) - h(t) = a q [N]_{q^2 }( t(q-1) + w )^2
\ee
Taking the limit $ N \rightarrow \infty $, the eq.(38) is then given by
\be
h(t) = h(w_0 ) + \frac { a q }{ q^2 -1 } ( t(q-1) + w )^2,
\ee
which gives the following :
\be
x(qt+w ) - q x(t) = h(w_0 ) + \frac { a q }{ q^2 -1 } ( t(q-1) + w )^2
\ee
This equation is still satisfied by replacing as follows :
\be
x(t) = \tilde{x} (t) + C t - C w_0 ,
\ee
where the constant $C$ will be fixed by demanding the initial condition later. The eq.(41) is then given by
\be
\tilde{x} (qt+w) -q \tilde{x} (t) = h(w_0 ) + \frac { a q }{ q^2 -1 } ( t(q-1) + w )^2
\ee
Iterating the eq.(42) $N$ times, we have
\be
q^{1-N} \tilde{x} (q^N t+[N]_q w )-q\tilde{x} (t) =  h(w_0 ) + \frac { a q }{ q^2 -1 } ( t(q-1) + w )^2
\ee
Inserting $t=w_0 $ into the eq.(36), we obtain
\be
h(w_0 ) = (1-q) x( w_0 ) ,
\ee
which enables us to replace the eq.(43) with the following equation by taking the limit $N \rightarrow \infty $ :
\be
\tilde{x} (t) = \tilde{x}(0) + \frac {a}{[2]_q } t^2 - \frac { 2 a w_0 }{[2]_q } t + \frac { a w_0^2 }{[2]_q }
\ee
Using $ \tilde{x}(0) = x(0) $ and the eq.(41), we have
\be
x(t) = x(w_0) -C w_0 +  \frac { a w_0^2 }{[2]_q } + \lb  C- \frac { 2 a w_0 }{[2]_q } \rb t + \frac {a}{[2]_q } t^2
\ee
Inserting $t=0$ into the eq.(46), we have
\be
x(0) = x(w_0) -C w_0 +  \frac { a w_0^2 }{[2]_q }
\ee
Applying the Hahn's derivative to the deformed position given in the eq.(46), we have
\be
v(t) = D_t x (t) = \lb  C- \frac { 2 a w_0 }{[2]_q } \rb  + a t
\ee
Inserting $t=0$ into the eq.(48), we obtain
\be
C = v(0) + \frac {2 a w_0 }{[2]_q }
\ee
Therefore the deformed position becomes
\be
x(t) = x(0) + v(0) t + \frac {1}{[2]_q } a t^2 ,
\ee
where
\be
x(0) = x(w_0) - v(0)  w_0 -  \frac { a w_0^2 }{[2]_q }
\ee

\section{ Vertical motion of a body in a resisting medium }

In the ($q,w$)-deformed mechanics, the ($q,w$)-deformed Newton's equation is easily derived from the eq.(2) as follows :
\be
F= m a = m D_t v ,
\ee
where  $v(t)$ and $a(t)$ are the q-deformed velocity and q-deformed acceleration, respectively and $m $ is a mass of the body.
When the force is constant, the body moves with the constant acceleration $ \frac {F}{m } $.
It should be emphasized that the force is not necessarily constant and, indeed, it may consist of several distinct parts.
First let us discuss the one dimensional motion with air resistance proportional to the ($q,w$)-deformed average velocity. Then the ($q,w$)-deformed equation of motion is given by
\be
m D_t v = -k \frac { v(t) + v(qt+w) }{[2]_q},
\ee
where $\frac { v(t) + v(qt+w) }{[2]_q}$ is called a ($q,w$)-deformed average velocity and becomes a instantaneous velocity when the deformation parameter $q$ goes to $1$ and $w$ goes to $0$.
The parameter $k$ is a positive constant that specifies the strength of the retarding force.
From the definition of the ($q,w$)-deformed time derivative, the eq.(53) has the following form :
\be
\lb 1- \frac { k  }{m[2]_q } ((q-1)t+w)   \rb v(t) = \lb 1+ \frac { k }{m[2]_q  } ((q-1)t+w) \rb v(qt+w)
\ee
By iterating this functional equation $N-1$ times, we obtain
\be
v(t) = \frac { \lb -\frac { k }{m[2]_q  } ((q-1)t+w) ; q \rb_N }{\lb  \frac { k }{m[2]_q  } ((q-1)t+w) ; q \rb_{N} } v(q^N t +[N]_q w),
\ee
where the q-shifted factorial is defined by
\be
(a;q)_0 =1, ~~~  (a;q)_N = (1-a)(1-q a)(1-q^2a ) \cdots (1-q^{N-1} a) , ~~~n=1, 2, 3, \cdots
\ee
Iterating the functional equation infinite times gives the following :
\be
v(t) = \frac { \lb -\frac { k }{m[2]_q  } ((q-1)t+w) ; q \rb_{\infty} }{\lb  \frac { k }{m[2]_q  } ((q-1)t+w) ; q \rb_{\infty} }v(w_0),
\ee
where we used the following relation :
\be
\lim_{N \rightarrow \infty } q^N = 0, ~~~\lim_{N \rightarrow \infty } [N]_q w = w_0
\ee

If the initial velocity $v(0)$ is set to $v_0$, we have
\be
v(w_0 ) = v_0
\ee
Then we have
\be
v(t) = v_0 \frac { \lb -\frac { k }{m[2]_q  } ((q-1)t+w) ; q \rb_{\infty} }{\lb  \frac { k }{m[2]_q  } ((q-1)t+w) ; q \rb_{\infty} } 
\ee

Now we introduce the well-known ($q,w$)-deformed exponential function satisfying
\be
D_t e_{q,w} (at) = a e_{q,w} (t),
\ee
where $a$ is an arbitrary constant. The ($q,w$)-deformed exponential can be written by using the q-shifted factorial as follows :
\be
e_{q,w} (at) = \frac {1}{ (- a((q-1)t+w) ;q)_{\infty} }
\ee
Inserting the eq.(62) into the eq.(60) gives the following expression :
\be
v(t) = v_0 \frac { e_{q,w} \lb - \frac{k}{m[2]_q } t \rb }{ e_{q,w} \lb  \frac{k}{m[2]_q } t \rb }
\ee
Indeed, we can easily check that the above expression reduce to the classical result
\be
v(t) = v_0 e^{ - k \frac {k}{m} t },
\ee
when the deformation parameter $q$ goes to $1$ and $w$ goes to $0$.

Now we study more complicated problem. We discuss the vertical motion of a body in a resisting medium in which there again exists a retarding force proportional to the ($q,w$)-deformed average velocity. Let us consider that the body is projected downward with zero initial velocity $v(0) =0$ in a uniform gravitational field. The ($q,w$)-deformed equation of motion is then given by
\be
m D_t v = m g -k \frac { v(t) + v(qt+w) }{[2]_q}
\ee

From the definition of the ($q,w$)-deformed time derivative, the eq.(65) has the following form :
\be
\lb 1- \frac { k}{m[2]_q  } ((q-1)t+w)  \rb v(t) = - g( (q-1)t +w) +  \lb 1+ \frac { k }{m[2]_q  } ((q-1)t +w) \rb v(qt+w)
\ee
By iterating this functional equation $N-1$ times, we obtain

\be
v(t) = \frac { \lb -\frac { k }{m[2]_q  } ((q-1)t+w) ; q \rb_N }{\lb  \frac { k }{m[2]_q  } ((q-1)t+w) ; q \rb_{N} } v(q^N t +[N]_q w )
-g ((q-1)t+w) \sum_{j=0}^{N-1}  \frac { q^j \lb \frac { k }{m[2]_q  } ((q-1)t+w) ; q \rb_j }{\lb  \frac { k}{m[2]_q  } ((q-1)t+w) ; q \rb_{j+1} }
\ee
The eq.(67) can be written as the following form :
\bd
v(t) = \frac { \lb -\frac { k }{m[2]_q  }((q-1) t+w) ; q \rb_N }{\lb  \frac { k }{m[2]_q  } ((q-1)t+w) ; q \rb_{N} } v(q^N t +[N]_q w  )
\ed
\be
+ \frac { [2]_q m g/k } {\lb  \frac { k }{m[2]_q  } ((q-1)t+w) ; q \rb_{N} } \sum_{n=0}^{[\frac {N-1}{2}]} \frac {(q;q)_N }{(q;q)_{2n+1} (q;q)_{N-1-2n} } q^{n(2n+1)} \lb \frac { k }{m[2]_q  } ((q-1)t+w) \rb^{2n+1},
\ee
where $[x] $ is a Gauss symbol.
Because the initial velocity $v(0)$ is zero and $\lim_{N \rightarrow \infty } q^N = 0$, we have following relation by taking the limit $ N \rightarrow \infty $ :

\bd
v(t) = \frac { \lb -\frac { k }{m[2]_q  } ((q-1)t+w) ; q \rb_{\infty} }{\lb  \frac { k }{m[2]_q  } ((q-1)t+w) ; q \rb_{\infty} } v(w_0)
\ed
\be
+\lim_{N \rightarrow \infty } \frac { [2]_q m g/k } {\lb  \frac { k }{m[2]_q  } ((q-1)t+w) ; q \rb_{N} } \sum_{n=0}^{[\frac {N-1}{2}]} \frac {(q;q)_N (1-q)^{2n+1} q^{n(2n+1)} }{(q;q)_{2n+1} (q;q)_{N-1-2n} }  \lb \frac { k }{m[2]_q  } (-(t-w_0) ) \rb^{2n+1} ,
\ee

Because $(q;q)_N $ and $(q;q)_{N-1-2n}$ reduce to $1$ when $N$ goes to the infinity, the expression (32) takes the following form :

\bd
v(t) =  \frac { \lb \frac { k }{m[2]_q  } ((q-1)t+w) ; q \rb_{\infty} }{\lb  \frac { k }{m[2]_q  } ((q-1)t+w) ; q \rb_{\infty} } v(w_0)
 \ed
 \be
+  \frac { [2]_q m g/k } {\lb  \frac { k }{m[2]_q  } ((q-1)t+w) ; q \rb_{\infty } } \sum_{n=0}^{\infty } \frac { (1-q)^{2n+1}  q^{n(2n+1)} }{(q;q)_{2n+1} } \lb \frac { k }{m[2]_q  } (-(t-w_0) )  \rb^{2n+1}
\ee

or
\bd
v(t) =  \frac { e_{q,w} \lb -\frac { k }{m[2]_q  }t \rb }{ e_{q,w} \lb  \frac { k }{m[2]_q  } t  \rb } v(w_0)
 \ed
\be
   +\frac { [2]_q m g}{k } e_{q,w} \lb - \frac { k }{m[2]_q  } t \rb  \sum_{n=0}^{\infty } \frac { (1-q)^{2n+1}  q^{n(2n+1)} }{(q;q)_{2n+1}  } \lb \frac { k }{m[2]_q  } (-(t-w_0) )  \rb^{2n+1}
\ee

The eq.(71) can be written by using the q-number factorial as follows :
\be
v(t) =  \frac { e_{q,w} \lb -\frac { k }{m[2]_q  }t \rb }{ e_{q,w} \lb  \frac { k }{m[2]_q  } t \rb } v(w_0)
+ \frac { [2]_q m g}{k } e_{q,w} \lb - \frac { k }{m[2]_q  } t \rb  \sum_{n=0}^{\infty } \frac {   q^{n(2n+1)} }{[2n+1]_q !  } \lb \frac { k }{m[2]_q  }  (-(t-w_0) ) \rb^{2n+1},
\ee
where the q-number factorial is defined by
\be
[n]_q ! = [n]_q [n-1]_q \cdots [2]_q [1]_q
\ee
Comparing the eq.(72) with the definition of the q-deformed exponential function such as
\be
e_q (x) = \sum_{n=0}^{\infty } \frac {   1 }{[n]_q !  }x^n ,
\ee
we can rewrite the eq.(72) as
\be
v(t) =  \frac { e_{q,w} \lb -\frac { k }{m[2]_q  }t \rb }{ e_{q,w} \lb  \frac { k }{m[2]_q  } t  \rb } v(w_0)
+  \frac { [2]_q m g}{k } e_{q,w} \lb - \frac { k }{m[2]_q  } t \rb  \sum_{n=0}^{\infty } \frac {   1}{ [2n+1]_{q^{-1}} !  } \lb \frac { k }{m[2]_q  } (-(t-w_0) ) \rb^{2n+1},
\ee
where we used the following relation
\be
[n]_{q^{-1}} ! = q^{\frac {-n(n-1)}{2}} [n]_q !~~~and ~~~[n]_{q^{-1}} = \frac { 1-q^{-n}}{1-q^{-1}}
\ee

Thus we can express the eq.(76) by using the q-deformed exponential function as follows :
\be
v(t) =  \frac { e_{q,w} \lb -\frac { k }{m[2]_q  }t \rb }{ e_{q,w} \lb  \frac { k }{m[2]_q  } t  \rb } v(w_0)
+ \frac { [2]_q m g}{2k }  e_{q,w} \lb - \frac { k }{m[2]_q  } t \rb
\left[  e_{q^{-1}} \lb  \frac { k }{m[2]_q  } t \rb -  e_{q^{-1}} \lb  -\frac {  k }{m[2]_q  } t \rb
  \right] ,
\ee
where we  used the following identity :
\be
e_{q^{-1}}(a) - e_{q^{-1}}(-a) =2 \sum_{n=0}^{\infty} \frac {a^{2n+1} } { [2n+1]_{q^{-1}} !}
\ee
Inserting $t=0$ into the eq.(77), we have 
\be
v(0) = v(w_0)
\ee
Then the equation (77) is written by 
\be
v(t) =  v(0) \frac { e_{q,w} \lb -\frac { k }{m[2]_q  }t \rb }{ e_{q,w} \lb  \frac { k }{m[2]_q  } t  \rb } 
+ \frac { [2]_q m g}{2k }  e_{q,w} \lb - \frac { k }{m[2]_q  } t \rb
\left[  e_{q^{-1}} \lb  \frac { k }{m[2]_q  } t \rb -  e_{q^{-1}} \lb  -\frac {  k }{m[2]_q  } t \rb
  \right] 
\ee
If the initial velocity is set to $0$, we have 
\be
v(t) =  \frac { [2]_q m g}{2k }  e_{q,w} \lb - \frac { k }{m[2]_q  } t \rb
\left[  e_{q^{-1}} \lb  \frac { k }{m[2]_q  } t \rb -  e_{q^{-1}} \lb  -\frac {  k }{m[2]_q  } t \rb
  \right] 
\ee
It can be easily checked that the expression (77) reduce to
\be
v(t) = \frac {m g}{k} \lb 1 - e^{ - kt/m } \rb, 
\ee
when he deformation parameter $q$ goes to $1$.

\section{Conclusion}

In this paper, we used the quantum variational calculus related to the Hahn's calculus [8] to construct the ($q,w$)-deformed version for the classical mechanics. We dealt with the motion with constant velocity and the motion with constant acceleration by using the first order ($q,w$)- difference equation. We also obtained the same result by using the second order ($q,w$)- difference equation  Moreover, we extended our work to the motion of a body in a resisting medium, where the retarding force is assumed to be proportional to the ($q,w$)-deformed average velocity. In this process, we found the new identity for infinite q-series.

We think that our work will be extended to more complicated mechanical problem such as ($q,w$)-deformed damped oscillator problem , ($q,w$)-deformed Kepler problem and so on. We hope that these topics will be solved in the near future.

%%%%%%%%%%%%%%%%%% References
%%%%%%%%%%%%%%%%%%%%%%%%%%%%%%%%%%%%%%%%%%%%%%%%%%%%%%%%%%%%%%%%%%%%%%%
\def\JMP #1 #2 #3 {J. Math. Phys {\bf#1},\ #2 (#3)}
\def\JP #1 #2 #3 {J. Phys. A {\bf#1},\ #2 (#3)}

%%%%%%%%%%%%%%%%%%%%%%%%%%%%%%%%%%%%%%%%%%%%%%%%%%%%%%%%%%%%%%%%%%%%%%%

\section*{Refernces}

[1] Almeida, R.,Torres,D.F.M., J.Math.Anal.Appl.{\bf 359}(2), 674 (2009)

[2] Bangerezako,G., J.Math.Anal.Appl.{\bf 289}(2), 650 (2004)

[3] Bangerezako,G., J.Math.Anal.Appl.{\bf 306}(1), 161 (2005)

[4] Cresson, J., Freserico, G.S.F., Torres,D.F.M., Topol. Methods Nonlinear Anal. {\bf 33}(2), 217 (2009)

[5] Kac, V., Cheung, P., {\it Quantum Calculus } , Springer, New York (2002)

[6] Jackson,F.H., Mess.Math. {\bf 38 }, 57 (1909).

[7] Hahn, W., Math.Nachr.{\bf 2 }, 4 (1949)

[8] Malinowska,A.B., Torres, D.F.M., math.OC 1006.3765v1 (2010)

\end{document}